\documentclass[paper,notoc]{JHEP3}
\usepackage{amsmath,amssymb}
\usepackage{graphicx}

\newcommand\fverb{\setbox\fverbbox=\hbox\bgroup\verb}
\newcommand\fverbdo{\egroup\medskip\noindent%
			\fbox{\unhbox\fverbbox}\ }
\newcommand\fverbit{\egroup\item[\fbox{\unhbox\fverbbox}]}
\newbox\fverbbox

\newcommand\EBox{\mathrm{Box}}
\newcommand\DBox{\mathrm{DBox}}
\newcommand\XBox{\mathrm{XBox}}
\newcommand\Hexa{\mathrm{Hexa}}
\newcommand\Tria{\mathrm{Tria}}
\newcommand\Hard{\mathrm{Hard}}
\newcommand\SHard{\mathrm{SHard}}

\newcommand\izero{i\varepsilon}
\newcommand\Dconf{d}

\title{Loop lessons from Wilson loops in ${\cal N}=4$ supersymmetric
  Yang-Mills theory}

\author{Charalampos Anastasiou, Andrea Banfi\\
  ETH Zurich, 8093 Zurich, Switzerland\\
  E-mail: \email{banfi@itp.phys.ethz.ch}}

\preprint{}

\abstract{ 
  ${\cal N}=4$ supersymmetric Yang-Mills theory exhibits a rather
  surprising duality of Wilson-loop vacuum expectation values and
  scattering amplitudes.  In this paper, we investigate this
  correspondence at the diagram level.  We find that one-loop
  triangles, one-loop boxes, and two-loop diagonal boxes can be cast
  as simple one- and two- parametric integrals over a single propagator
  in configuration space. We observe that the two-loop Wilson-loop
  ``hard-diagram'' corresponds to a four-loop hexagon Feynman
  diagram.  Guided by the diagrammatic correspondence of the
  configuration-space propagator and loop Feynman diagrams, we derive
  Feynman parameterizations of complicated planar and non-planar
  Feynman diagrams which simplify their evaluation. For illustration,
  we compute numerically a four-loop hexagon scalar Feynman diagram.
}


\begin{document}

\section{Introduction}
\label{sec:intro}

${\cal N}=4$ supersymmetric Yang-Mills theory is rich of symmetries
and surprising dualities. In the late 90s, Maldacena proposed that this
conformal field theory in flat space-time is dual in its planar limit
to a string theory in anti-de Sitter
space~\cite{Maldacena:1997re}. Later investigations showed that
anomalous dimensions of operators can be mapped to integrable quantum
mechanical systems of spin-chains~\cite{Minahan:2002ve,
  Beisert:2003ea, Beisert:2005fw}.

A surprising discovery was that the two-loop four-point planar
amplitude could be expressed entirely in terms of the corresponding
one-loop amplitude~\cite{Anastasiou:2003kj}. From the collinear limit
of gauge theory amplitudes it was conjectured that this factorization
may hold for all two-loop planar MHV amplitudes, while the AdS/CFT
correspondence was inviting a general factorization at all orders in
perturbation theory~\cite{Anastasiou:2003kj}.  A factorization of the
three-loop amplitude was proven in~\cite{Bern:2005iz} where an
explicit factorization ansatz valid at all orders was also
formulated. Two-loop factorization was shown to hold for the
five-point planar MHV amplitude~\cite{Cachazo:2006tj,Bern:2006vw}, 
however it was shown to break down for six-point scattering
amplitudes~\cite{Bern:2008ap}.
  
Alday and Maldacena exploited the AdS/CFT correspondence to evaluate
scattering amplitudes of ${\cal N}=4$ supersymmetric Yang-Mills theory
in the strong coupling limit, as a minimal surface on $\mathrm{AdS}_5$
bounded by a polygon with sides the momenta of the external
states~\cite{Alday:2007hr}. This has lead to the conjecture that
scattering amplitudes are dual to the vacuum expectation value of
Wilson-loops order by order in perturbation 
theory~\cite{Drummond:2007aua,Brandhuber:2007yx}.  
This duality has been tested with explicit comparisons at two loops for up to six-point
amplitudes and hexagon Wilson
loops~\cite{Bern:2008ap,Drummond:2008aq,Brandhuber:2010bj}.

The evaluation of Wilson-loop vacuum expectation values turns out to
be much simpler than the corresponding two-loop amplitudes.  Two-loop
Wilson loops are known analytically up to
hexagons~\cite{DelDuca:2009au,DelDuca:2010zg}, and numerically for an
arbitrary number of sides~\cite{Anastasiou:2009kn}.  The relative
simplicity of two-loop Wilson loops is due to that their Feynman
representations require at most five integration variables
irrespective of the number of the polygon sides. On the contrary, the
number of Feynman parameters for an amplitude increases with the
number of external legs.

${\cal N}=4$ supersymmetric Yang-Mills theory is rich of symmetries
which constrain highly the structure of scattering amplitudes. 
The amplitude/Wilson-loop duality may be attributed to these symmetries. 
Nevertheless, these amplitudes require the computation of highly complicated scalar
integrals which are typically the most complicated ``master''
integrals entering the evaluation of less symmetric amplitudes in
theories such as QCD. For example, all four-point planar
amplitudes in ${\cal N}=4$ supersymmetric Yang-Mills through two loops 
are simple expressions in terms of the one-loop and two-loop box scalar (master)
integrals~\cite{Green:1982sw,Bern:1997nh}.  These very amplitudes being dual
to a Wilson loop is not only a ``magic'' property of the theory but
also a remarkable property of the one-loop box master integral.

In this paper we show that there is a correspondence between individual 
Wilson-loop  diagrams  and usual Feynman diagrams.  We note that a
scalar one-loop triangle is dual to a propagator of the Wilson-loop
configuration space which joins a fixed point and a line segment.  We
also find that the one-loop box in six dimensions and the two-loop
diagonal-box in four dimensions with two light-like non-adjacent legs
are dual to a configuration-space propagator joining two
line segments.

An intriguing Wilson-loop two-loop diagram is the so called ``hard
diagram''. It consists of a triple-gluon vertex connected via gluon
propagators to three sides of a polygon Wilson-loop.  An analytic
solution for this diagram is fully known for square and pentagon
Wilson loops~\cite{Drummond:2007aua,Drummond:2007au} while it is known
in Regge kinematics\footnote{This limit is sufficient for the exact
  determination of the two-loop hexagon Wilson-loop for arbitrary
  kinematics.}  for a hexagon
Wilson-loop~\cite{DelDuca:2009au,DelDuca:2010zg} and  for an octagon in 
special kinematic configurations~\cite{DelDuca:2010zp}.  
The ``hard diagram'' has only been computed numerically with standard methods for
polygon Wilson loops with more than six
sides~\cite{Anastasiou:2009kn}.  In this paper, we demonstrate that
the Wilson-loop ``hard diagram'' is dual to a four-loop hexagon
diagram with four one-loop triangle subgraphs.  We also derive a
representation of the propagator in configuration space connecting a
fixed point and a line segment as the product of two massive
propagators.  With this representation, we find that the ``hard
diagram'' is dual to a one-loop hexagon massive Feynman integral
integrated over its own mass parameters.

We hope that these representations of the ``hard diagram'' with an
arbitrary number of polygon sides will become a convenient starting
point for an analytic evaluation of it in the future.  However, this
and other two-loop Wilson-loop diagrams are relatively easy to
evaluate numerically~\cite{Anastasiou:2009kn}.  We can then exploit
the diagrammatic dualities of Wilson loops and amplitudes to
facilitate the computation of complicated Feynman integrals in
amplitudes. For illustration, we present here a numerical evaluation
of a scalar four-loop hexagon integral with light-like legs which is
mapped to a scalar hexagon Wilson-loop ``hard diagram'' times a
$1/\epsilon^3$ prefactor.

Inspired by the diagrammatic dualities we have found, we can derive
simple representations for multi-loop integrals with ``easy-box''
subgraphs (boxes with two non-adjacent light-like legs). These are
non-planar diagrams which are rather cumbersome to evaluate naively.
We replace ``easy-box'' subgraphs by a single propagator reducing the
number of loops by one.  All two-loop non-planar integrals with such a
subgraph are reduced to one-loop integrals where two of the external
momenta are variables constructed as linear combinations of the only
two external momenta entering the ``easy-box'' subgraph.  A two-fold
integration is also required over a range of such linear combinations
for the external momenta.  These representations require a smaller
number of integration variables than canonical Feynman
parameterizations.  They also lead to simple Mellin-Barnes
representations.

Our article is organized as follows. In
Section~\ref{sec:triangle-1loop} we derive the correspondence between
a one-loop triangle and a propagator in configuration space.  In
Section~\ref{sec:planar-boxes} this correspondence is exploited to
rewrite planar easy boxes as a configuration-space propagator joining
two line segments.  In Section~\ref{sec:hexa-4loop} we consider the
``hard diagram'' contribution to a two-loop Wilson loop and show that
it can be mapped into a four-loop hexagon. This correspondence is then
used for a numerical evaluation of the four-loop hexagon, which we
perform in Section~\ref{sec:hexa-num}. In
Section~\ref{sec:triangle-mass} we derive an alternative
representation of the configuration-space propagator, and use it to
make the four-loop hexagon of Section~\ref{sec:hexa-4loop} correspond
to a one-loop hexagon with massive internal propagators, integrated
over the internal masses. We conclude in Section~\ref{sec:non-planar}
by applying the correspondence we have found for easy boxes to obtain
better Feynman parameterizations for non-planar diagrams.

\section{A space-time propagator as a one-loop triangle}
\label{sec:triangle-1loop} 

\begin{figure}[h]
\begin{center}
\includegraphics[height=3cm]{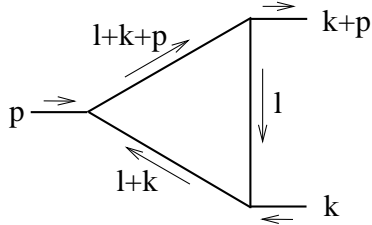}
\end{center}
\caption{A one-loop triangle with a massless leg $p$.}
\label{fig:triangle}
\end{figure}

\begin{figure}[h]
\begin{center}
  \includegraphics{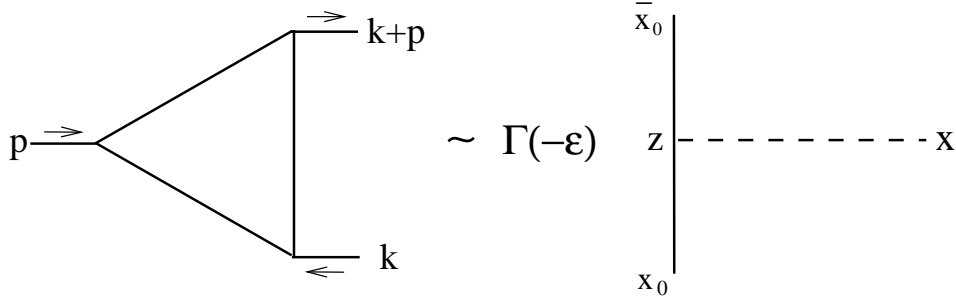}
\end{center}
  \caption{Pictorial representation of the mapping of the triangle
    into a Wilson line.}
  \label{fig:scalar-line}
\end{figure}

The basic object entering the evaluation of the vacuum expectation
values of a Wilson loop is a scalar propagator in configuration
space. For a massless scalar theory in $D = 4-2\epsilon_{UV}$
dimensions, this is
\begin{equation}
  \label{eq:scalar-prop}
  \Delta(x)\equiv i\!\! \int \frac{d^Dk}{(2\pi)^{D}} 
  \frac{e^{-i k x}}{k^2+\izero}=
\frac{1}{4\pi^{\frac{D}{2}}} 
  \frac{\Gamma\left(\frac{D}{2}-1\right)}
  {(-x^2+\izero)^{\frac{D}{2}-1}} = 
\frac{1}{4\pi^{2-\epsilon_{UV}}} \frac{\Gamma(1-\epsilon_{UV})}
{(-x^2+\izero)^{1-\epsilon_{UV}}}\,.
\end{equation}
There is  a correspondence between this propagator and a scalar
Feynman diagram. Consider a one-loop triangle with massless internal propagators 
(Fig.~\ref{fig:triangle}) 
\begin{equation}
  \label{eq:triangle-1loop}
  \Tria(k,p;1,1,1) = \int \frac{d^D \ell}{i\pi^{\frac{D}{2}}}
  \frac{1}{(\ell^2+\izero)\,[(\ell+k)^2+\izero]\,[(\ell+k+p)^2+\izero]}\,
\end{equation}
with
\begin{equation}
p^2=0, \quad k^2,(k+p)^2\neq 0. 
\end{equation}
We combine with a Feynman parameter the last two propagators of the above expression, 
\begin{equation}
  \label{eq:triangle-bubble}
  \Tria(k,p;1,1,1)  = \int_0^1 d\tau \!\!\int \frac{d^D \ell}{i\pi^{\frac{D}{2}}}
  \frac{1}{[\ell^2+\izero]\,[(\ell+k+\tau p)^2+\izero]^2}, 
\end{equation}
and integrate the loop momentum $\ell$ in $D=4-2\epsilon$ dimensions. 
We obtain: 
\begin{equation}
  \label{eq:triangle-tau}
  \begin{split}
    \Tria(k,p;1,1,1) &= -\Gamma(-\epsilon)\frac{\Gamma(1-\epsilon)
      \Gamma(1+\epsilon)}{\Gamma(1-2\epsilon)}
    \int_0^1 
    \frac{d\tau}{[-(k+\tau p)^2-\izero]^{1+\epsilon}}\,.
  \end{split}
\end{equation}
The propagator in the above equation resembles already the
configuration space propagator in Eq.~\eqref{eq:scalar-prop}. To make
the correspondence explicit we introduce a ``space-time
point'' $x=k$ and a ``trajectory'' in configuration space
$z(\tau)=-\tau p$.  By comparing Eq.~(\ref{eq:triangle-tau}) with the
propagator in Eq.~(\ref{eq:scalar-prop}), and identifying 
$\epsilon_{UV}=-\epsilon$, we find
\begin{equation}
  \label{eq:x-triangle}
\Tria(k,p;1,1,1)=  -4\pi^{2+\epsilon}
  \frac{\Gamma(-\epsilon)\Gamma(1-\epsilon)}{\Gamma(1-2\epsilon)}
  \int_0^1 \!d\tau \,\Delta^*(x-z(\tau))\,.
\end{equation}
$\Delta^*(x)$ is the complex conjugate of the propagator
$\Delta(x)$ of Eq.  (\ref{eq:scalar-prop}), i.e.\ is an anti-causal
propagator, appearing when evaluating conjugate amplitudes.
This correspondence is illustrated in
Fig.~\ref{fig:scalar-line}.  
We notice that the result in Eq.~\eqref{eq:x-triangle} is invariant
under translations, so that we can equivalently define $x=x_0+k$ and
$z(\tau)=x_0-\tau p$, where $x_0$ is an arbitrary space-time point.

\subsection{The triangle rule}
\label{sec:triangle}

\begin{figure}
\begin{center}
\includegraphics{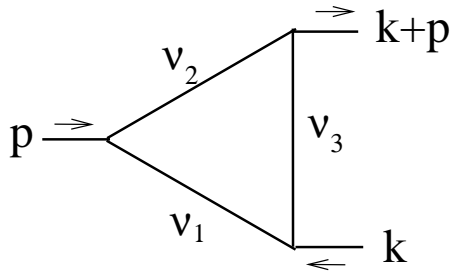}
  \caption{A scalar triangle with arbitrary powers in the propagators.}
  \label{fig:triangle-powers}
\end{center}
\end{figure}

We now consider a  one-loop triangle (Fig.~\ref{fig:triangle-powers}) 
with general powers of propagators 
\begin{equation}
  \label{eq:tria-arbitrary}
  \Tria(k,p;\nu_1,\nu_2,\nu_3) = \int \frac{d^D\ell}{i \pi^{\frac{D}{2}}}
  \frac{1}{[\ell^2]^{\nu_1} [(\ell+p)^2]^{\nu_2}[(\ell-k)^2]^{\nu_3}}\,.
\end{equation}
We implicitly assume that all denominators are regularized by giving
each of them a positive infinitesimal imaginary part. 
Following the same procedure as above, we derive:
\begin{eqnarray}
  \label{eq:tria-final}
  \Tria(k,p,\nu_1,\nu_2,\nu_3) &=& (-1)^{\frac{D}{2}} 
  \frac{\Gamma\left(\nu_{123}-\frac{D}{2}\right)}
  {\Gamma(\nu_1)\Gamma(\nu_2)\Gamma(\nu_3)}
 \frac{\Gamma\left(\frac{D}{2}-\nu_{12}\right)
   \Gamma\left(\frac{D}{2}-\nu_{3}\right)}{\Gamma(D-\nu_{123})}
\nonumber \\
&& \times
  \int_0^1\!\! d\tau 
  \frac{(1- \tau)^{\nu_1-1} \tau^{\nu_2-1}}{[(k+\tau
    p)^2]^{\nu_{123}-\frac{D}{2}}} \,.
\end{eqnarray}
We notice that, unless $\nu_1=\nu_2=1$, the interpretation of the
one-loop triangle as a Wilson-line propagator is lost due to the non-trivial 
numerator of the integrand. For $\nu_1=\nu_2=1$, the one-loop triangle can be interpreted
as a propagator attached to a Wilson line, since
Eq.~\eqref{eq:tria-final} can be rewritten as follows:
\begin{multline}
  \label{eq:tria-final-11}
  \Tria(k,p,1,1,\nu_3) = (-1)^{\nu_3} 
  \frac{\Gamma\left(\nu_3+\epsilon\right)}
  {\Gamma(\nu_3)}
  \times \\ \times
 \frac{\Gamma\left(-\epsilon\right)
    \Gamma\left(2-\epsilon-\nu_{3}\right)}{\Gamma(2-2\epsilon-\nu_3)}
  \int_0^1\!\! d\tau 
  \frac{1}{[-(k+\tau p)^2]^{\nu_3+\epsilon}} \,,
\end{multline}
where we have set $D=4-2\epsilon$. Comparing with the scalar
propagator in Eq.~\eqref{eq:scalar-prop}, we find the correspondence
\begin{multline}
  \label{eq:tria-11-scalprop}
  \Tria(k,p,1,1,\nu_3) = (-1)^{\nu_3} 
 \frac{4\pi^{1+\epsilon+\nu_3}\Gamma\left(-\epsilon\right)}{\Gamma(\nu_3)}
  \frac{\Gamma\left(2-\epsilon-\nu_{3}\right)}{\Gamma(2-2\epsilon-\nu_3)}
  \times \\ \times \int_0^1\!\! d\tau
  \left.\Delta^*(x-z(\tau))\right|_{\epsilon_{UV}=-\epsilon-\nu_3+1}\,,
\end{multline}
where, as before, we have identified $k$ with the space-time point $x$
and have introduced the trajectory $z(\tau) = -\tau p$.  Therefore a 
one-loop triangle $\Tria(k,p,1,1,\nu_3)$ in $D=4-2\epsilon$ is dual to
the propagator in configuration space  in $D=4+2 \nu_3 +  2 \epsilon$ dimensions.  
We remark that the duality of the propagator in configuration space and the one-loop triangle has  
been also discussed in Refs.~\cite{Gorsky:2009nv,Gorsky:2009dr}.\footnote{We  thank Gregory Korchemsky for 
bringing these articles  to our  attention}

\section{The two-loop diagonal box as a one-loop Wilson loop diagram}
\label{sec:planar-boxes}

\begin{figure}[h]
\begin{center}
\includegraphics[width=.5\textwidth]{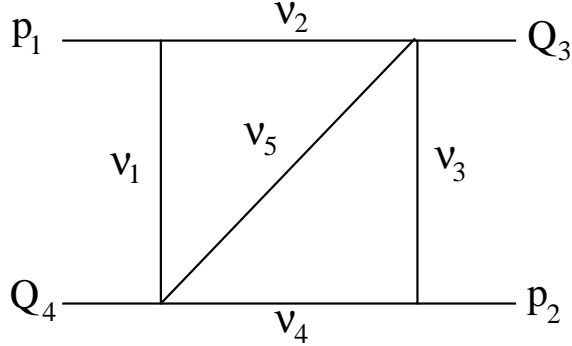}
\end{center}
  \caption{The diagonal box with massless internal propagators raised to
    arbitrary powers $\nu_i$.}
\label{fig:diagonal-box}
\end{figure}

A  question which arises from the duality of the previous  section
is whether similar dualities exist for more complicated diagrams.
We  shall show  that the two-loop diagonal box  diagram  in
Fig.~\ref{fig:diagonal-box} is dual to a one-loop Wilson-loop
diagram.  
The two-loop diagonal box is
\begin{eqnarray}
  \label{eq:dbox}
 && \DBox(p_1,p_2,Q_3;\nu_1,\nu_2,\nu_3,\nu_4,\nu_5) = \nonumber \\
 && \quad 
 \int \frac{d^D k d^D\ell}{i^2\pi^D} 
  \frac{1}{[k_1^2]^{\nu_1}[(k_1+p_1)^2]^{\nu_2}[k_2^2]^{\nu_3}[(k_2+p_2)^2]^{\nu_4}
    [\ell^2]^{\nu_5}}\,,
\end{eqnarray}
where $k_1=k+\ell$ and $k_2=k+p_1+Q_3$.  We can cast this integral as
\begin{equation}
  \label{eq:dbox-triangle}
  \DBox(p_1,p_2,Q_3;\nu_1,\nu_2,\nu_3,\nu_4,\nu_5) = 
  \int \frac{d^D k}{i\pi^{\frac{D}{2}}} 
  \frac{\Tria(k,p,\nu_1,\nu_2,\nu_5)}{[k_2^2]^{\nu_3}[(k_2+p_2)^2]^{\nu_4}}\,.
\end{equation}
and then use the triangle rule of Eq.~(\ref{eq:tria-final}) on 
$\Tria(k,p,\nu_1,\nu_2,\nu_5)$. We obtain
\begin{multline}
  \label{eq:dbox-triangle-done}
  \DBox(p_1,p_2,Q_3;\nu_1,\nu_2,\nu_3,\nu_4) = 
  (-1)^{\frac{D}{2}} 
  \frac{\Gamma\left(\nu_{125}-\frac{D}{2}\right)}
  {\Gamma(\nu_1)\Gamma(\nu_2)\Gamma(\nu_5)}
 \frac{\Gamma\left(\frac{D}{2}-\nu_{12}\right)
    \Gamma\left(\frac{D}{2}-\nu_{5}\right)}{\Gamma(D-\nu_{125})}
 \times  \\ 
 \times
  \int_0^1\!\! d\tau_1 
  \int \frac{d^D k}{i\pi^{\frac{D}{2}}} 
  \frac{( 1-\tau_1)^{\nu_1-1} \tau_1^{\nu_2-1}}
  {[(k+\tau_1 p_1)^2]^{\nu_{125}-\frac{D}{2}}
    [k_2^2]^{\nu_3}[(k_2+p_2)^2]^{\nu_4}}\,.
\end{multline}
We recognize that the $k$ integral is another triangle
\begin{equation}
  \label{eq:second-triangle}
  \int \frac{d^D k}{i\pi^{\frac{D}{2}}} 
  \frac{1}
  {[(k+\tau_1 p_1)^2]^{\nu_{125}-\frac{D}{2}}
    [k_2^2]^{\nu_3}[(k_2+p_2)^2]^{\nu_4}} = 
  \Tria\left(Q_3+(1-\tau_1) p_1,p_2,\nu_3,\nu_4,\nu_{125}-\frac D2\right)\,,
\end{equation}
and  apply again the triangle rule.  Our representation of the two-loop diagonal box reads
\begin{multline}
  \label{eq:dbox-final}
  \DBox(p_1,p_2,Q_3;\nu_1,\nu_2,\nu_3,\nu_4,\nu_5) = (-1)^{D}
  \frac{\Gamma\left(\nu_{12345}-D\right)}{\Gamma(\nu_1)\Gamma(\nu_2)\Gamma(\nu_3)\Gamma(\nu_4)\Gamma(\nu_5)}
  \times \\ \times
  \frac{\Gamma\left(\frac{D}{2}-\nu_{12}\right)
    \Gamma\left(\frac{D}{2}-\nu_{34}\right)
    \Gamma\left(\frac{D}{2}-\nu_{5}\right)}
  {\Gamma\left(\frac{3}{2}D-\nu_{12345}\right)}
  \times \\ \times
  \int_0^1 \! d\tau_1 \int_0^1 \! d\tau_2
  \,\frac{(1-\tau_1)^{\nu_1-1} \tau_1^{\nu_2-1} (1- \tau_2)^{\nu_3-1} \tau_2^{\nu_4-1}}
  {[((1- \tau_1) p_1+\tau_2 p_2+Q_3)^2]^{\nu_{12345}-D}}\,.
\end{multline}
\begin{figure}[h]
\begin{center}
\includegraphics[width=0.9\textwidth]{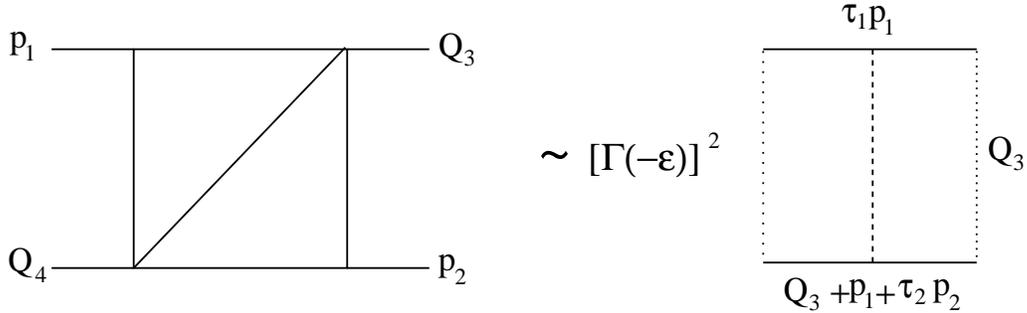}
\end{center}
  \caption{Schematic illustration of the correspondence between the
    diagonal box and a one-loop scalar Wilson loop. 
}
  \label{fig:diagonal-1loop}
\end{figure}
In the special case $\nu_1=\nu_2=\nu_3=\nu_4=1$,
\begin{multline}
  \label{eq:dbox-nui1}
  \DBox(p_1,p_2,Q_3,1,1,1,1,\nu_5) = 
  (-1)^{\nu_5}\frac{\Gamma\left(\nu_5+2\epsilon\right)\Gamma(2-\nu_5-\epsilon)\Gamma^2(-\epsilon)}
  {\Gamma(2-\nu_5-3\epsilon)\Gamma(\nu_5)}
  \times \\ \times
  \int_0^1 \! d\tau_1 \int_0^1 \! d\tau_2
  \,\frac{1}
  {[-(\bar \tau_1 p_1+\tau_2 p_2+Q_3)^2]^{\nu_5+2\epsilon}}\,.
\end{multline}
We can interpret the two-loop diagonal box as one of the one-loop
contributions to the vacuum expectation value of a Wilson loop with an
arbitrary number of edges (see Fig.~\ref{fig:diagonal-1loop}). This is
accomplished by introducing the trajectories $z(\tau_i) = x_i-\tau_i
\,p_i$, ending in the points $\bar x_i$ such that $p_i=x_i-\bar
x_i$. In terms of the $x_i$ and $\bar x_i$ the arbitrary momenta $Q_3$
and $Q_4$ are given by:
\begin{equation}
  \label{eq:Qi-xi-box}
  Q_3 = \bar x_1-x_2\,, \qquad   
  Q_4 = \bar x_2-x_1 \,.
\end{equation}
The exact correspondence is
\begin{multline}
  \label{eq:dbox-loop}
  \DBox(p_1,p_2,Q_3;1,1,1,1,\nu_5)=
  (-1)^{\nu_5}\, 4 \pi^{1+\nu_5+2\epsilon}\frac{\Gamma(2-\nu_5-\epsilon)\Gamma^2(-\epsilon)}
  {\Gamma(2-\nu_5-3\epsilon)\Gamma(\nu_5)}
  \times \\ \times
 \int_0^1 \! d\tau_1 \int_0^1 \! d\tau_2
  \>
  \Delta^*(z(\tau_1)-z(\tau_2))|_{\epsilon_{UV}=1-\nu_5-2\epsilon}\,.
\end{multline}
We further notice that for $Q_3=0$ one recovers the cusp contribution
to the  one-loop Wilson loop.

\subsection{The two-loop diagonal box as a one-loop easy box}
\label{sec:easy-box}

It is known that the two-loop diagonal box of the previous section 
and the one-loop easy box  are  dual.\footnote{We  thank Bas Tausk
  for bringing this duality to our attention in private discussions  immediately after
his publication of Ref.~\cite{Tausk:1999vh},  and Lance Dixon for presenting
recently to us an independent derivation.}  This  can be seen  easily
by comparing their Mellin-Barnes  representations~\cite{Tausk:1999vh}.  
This duality is easy to prove by showing  that the one-loop box (in
$D=6-4\epsilon$ dimensions)  and the two-loop diagonal box (in  $D=4-2\epsilon$
dimensions)  correspond to the same  Wilson-loop diagram of the
right-hand side of Eq.~(\ref{eq:dbox-loop}). 

\begin{figure}[h]
\begin{center}
\includegraphics[width=.5\textwidth]{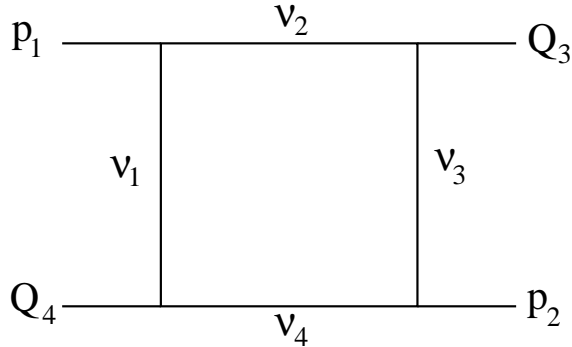}
\end{center}
\caption{The easy box.}
\label{fig:easy-box}
\end{figure}
The one-loop ``easy box'' of Fig.~\ref{fig:easy-box} is defined as:
\begin{equation}
  \label{eq:ebox}
  \EBox(p_1,p_2,Q_3,\nu_1,\nu_2,\nu_3,\nu_4) = 
  \int \frac{d^D k}{i\pi^{\frac{D}{2}}} 
  \frac{1}{[k_1^2]^{\nu_1}[(k_1+p_1)^2]^{\nu_2}[k_2^2]^{\nu_3}[(k_2+p_2)^2]^{\nu_4}}\,,
\end{equation}
with $k_1=k$, $k_2=k+p_1+Q_3$ and  $p_1^2=p_2^2=0$.  
We perform  the Feynman parameterization found in Ref.~\cite{Kramer:1986sr}.
We first join the two lines separated by $p_1$ by introducing the Feynman
parameter $\tau_1$
\begin{equation}
  \label{eq:ebox-triangle}
  \begin{split}
    \EBox(p_1,p_2,Q_3,\nu_1,\nu_2,\nu_3,\nu_4) &= 
    \frac{\Gamma(\nu_{12})}{\Gamma(\nu_1)\Gamma(\nu_2)}
    \int_0^1 d\tau_1 \!\!
    \int \frac{d^D k}{i\pi^{\frac{D}{2}}} 
    \frac{\bar \tau_1^{\nu_1-1} \tau_1^{\nu_2-1}}
    {[(k_1+\tau_1
      p_1)^2]^{\nu_{12}}[k_2^2]^{\nu_3}[(k_2+p_2)^2]^{\nu_4}}
\\ &=
    \frac{\Gamma(\nu_{12})}{\Gamma(\nu_1)\Gamma(\nu_2)}
    \int_0^1 d\tau_1 \,\bar \tau_1^{\nu_1-1} \tau_1^{\nu_2-1}\,
    \Tria(Q_3+\bar \tau_1 p_1,p_2;\nu_3,\nu_4,\nu_{12})
    \,.
  \end{split}
\end{equation}
In the above, we used the shorthand  notation 
$\bar \tau_i \equiv 1 - \tau_i$,
and we recognized that the $k$ integral corresponds to a
triangle. Applying the triangle rule
Eq.~(\ref{eq:tria-final}), we get
\begin{multline}
  \label{eq:ebox-final}
  \EBox(p_1,p_2,Q_3,\nu_1,\nu_2,\nu_3,\nu_4) = 
  (-1)^{\frac{D}{2}}
  \frac{\Gamma\left(\nu_{1234}-\frac{D}{2}\right)}
  {\Gamma(\nu_1)\Gamma(\nu_2)\Gamma(\nu_3)\Gamma(\nu_4)}
  \times \\ \times
 \frac{\Gamma\left(\frac{D}{2}-\nu_{12}\right)
    \Gamma\left(\frac{D}{2}-\nu_{34}\right)}
  {\Gamma\left(D-\nu_{1234}\right)}
  \int_0^1 \! d\tau_1 \int_0^1 \! d\tau_2
  \,\frac{\bar \tau_1^{\nu_1-1} \tau_1^{\nu_2-1} \bar \tau_2^{\nu_3-1} \tau_2^{\nu_4-1}}
  {[(\bar \tau_1 p_1+\tau_2 p_2+Q_3)^2]^{\nu_{1234}-\frac{D}{2}}}\,.
\end{multline}
Comparing Eq.~\eqref{eq:ebox-final} with Eq.~\eqref{eq:dbox-final} we immediately obtain that the diagonal box
is proportional to the one-loop easy box in $2(D-\nu_5)$
dimensions. The precise relation is
\begin{multline}
  \label{eq:ebox-dbox}
  \EBox_{2(D-\nu_5)}(p_1,p_2,Q_3,\nu_1,\nu_2\nu_3,\nu_4) =   
  (-1)^{-\nu_5} 
  \frac{\Gamma\left(\frac{3}{2}D-\nu_{12345}\right)}
  {\Gamma(2D-\nu_{12345}-\nu_5)}
  \times \\ \times
  \frac{\Gamma(D-\nu_{125})}{\Gamma\left(\frac{D}{2}-\nu_{12}\right)}
  \frac{\Gamma(D-\nu_{345})}{\Gamma\left(\frac{D}{2}-\nu_{34}\right)}
  \frac{\Gamma(\nu_5)}{\Gamma\left(\frac{D}{2}-\nu_{5}\right)}
  \times \\ \times
  \DBox_D(p_1,p_2,Q_3,\nu_1,\nu_2,\nu_3,\nu_4,\nu_5)  \,.
\end{multline} 

The connection to a Wilson-loop diagram is exact if  we  specialize Eq.~(\ref{eq:ebox-final}) to the case $\nu_i=1$. 
Replacing $D=4-2\epsilon$ we obtain
\begin{multline}
  \label{eq:ebox-nui1}
  \EBox(p_1,p_2,Q_3,1,1,1,1) = 
  2\,\Gamma(-\epsilon)\frac{\Gamma\left(2+\epsilon\right)\Gamma(1-\epsilon)}
  {\Gamma(1-2\epsilon)}
  \times \\ \times
  \int_0^1 \! d\tau_1 \int_0^1 \! d\tau_2
  \,\frac{1}
  {[-(\bar \tau_1 p_1+\tau_2 p_2+Q_3)^2]^{2+\epsilon}}\,.
\end{multline}
and, equivalently, 
\begin{multline}
  \label{eq:ebox-nui1-prop}
  \EBox(p_1,p_2,Q_3,1,1,1,1) = 
  8\pi^{3+\epsilon}\,\Gamma(-\epsilon)\frac{\Gamma(1-\epsilon)}
  {\Gamma(1-2\epsilon)} 
  \times \\ \times
  \int_0^1 \! d\tau_1 \int_0^1 \! d\tau_2
  \, \Delta^*(z(\tau_1)-z(\tau_2))|_{\epsilon_{UV}=-1-\epsilon}\,,
\end{multline}
with the trajectories $z(\tau_i)$ being the same as in
Fig.~\ref{fig:diagonal-1loop}.  Therefore, the one-loop easy box can
be interpreted as a one-loop diagram contributing to a Wilson loop
with an arbitrary number of edges in $D=6+2\epsilon$
dimensions. Equivalently, the same Wilson-loop diagram in
$D=4+2\epsilon$ dimensions corresponds to the one-loop easy box
diagram in $D=6-4\epsilon$ dimensions (which is finite).

\section{The hard diagram as a four-loop hexagon}
\label{sec:hexa-4loop}
\begin{figure}[h]
\begin{center}
 \includegraphics[width=.5\textwidth]{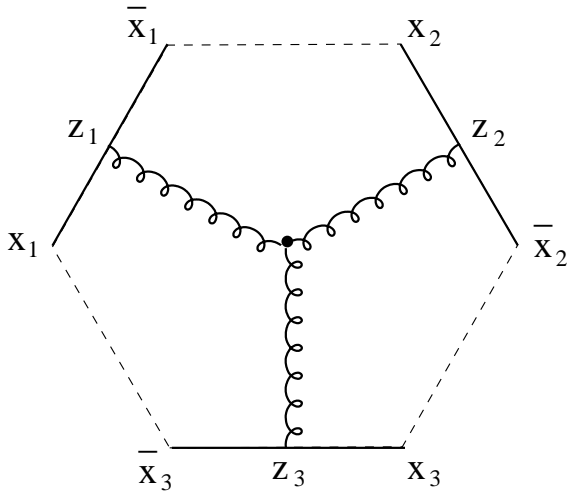}
\end{center}
  \caption{The hard diagram.}
  \label{fig:hard}
\end{figure}

In this section we seek further dualities  among Wilson loop diagrams
and  Feynman diagrams.  It has been shown by explicit calculations 
that the Wilson-loop and  loop-amplitude duality  holds for   two-loop 
six-leg amplitudes.  In the previous section we  saw  that one-loop
Wilson-loop diagrams are mapped  naturally to two-loop Feynman
diagrams and also,  with a handy 
duality of the two-loop diagonal box and the one-loop ``easy box'', 
to one-loop Feynman diagrams. 

We  now  examine the  so-called  ``hard diagram''  which 
contributes  to the two-loop Wilson loop.   
It owes  its name to the fact that  this has been very difficult to
compute analytically for polygon Wilson loops with more than five sides.
We shall show  that the hard diagram is in fact a four-loop
hexagon in disguise.  We will first show that the hard
diagram can be obtained by acting with a suitable differential
operator on a ``scalar hard diagram'', a special two-loop diagram
in configuration space. Then, we will use the triangle rule 
to map the scalar hard diagram into a four-loop hexagon.

The hard diagram is represented in Fig.~\ref{fig:hard}. There we
recognize three trajectories $z(\tau_i)$ for $i=1,2,3$,
which depend on the three light-like momenta $\{p_1,p_2,p_3\}$ via the
parameterization
\begin{equation}
  \label{eq:movin-points}
  z(\tau_i)= x_i-\tau_i\, p_i \equiv z_i \,.
\end{equation}
The other three positions $\{\bar x_1, \bar x_2, \bar x_3\}$ are
related to three momenta $\{Q_1,Q_2,Q_3\}$, which can be off-shell or
on-shell, as follows:
\begin{equation}
  \label{eq:xbari-Qi}
  \bar x_1 - x_2 = Q_3\,,\quad 
  \bar x_2 - x_3 = Q_1\,,\quad
  \bar x_3 - x_1 = Q_2\,.
\end{equation}
The above distances represent the insertion of an arbitrary number of
sides which are not connected with gluon propagators. The minimum
number of sides for which all $Q_i$ are non-zero is six, in which case
$\{Q_1,Q_2,Q_3\}$ are light-like.

Neglecting colour and symmetry factors, the explicit expression for
the hard diagram reads
\begin{eqnarray}
  \label{eq:hard-contracted}
 && \Hard(p_1,p_2,p_3,Q_1,Q_2,Q_3) = p_1^{\mu_1} p_2^{\mu_2}
  p_3^{\mu_3} \int_0^1 \left(\prod_{i=1}^3 d\tau_i\right)
 \nonumber    \\ 
&& \quad \times
  V^{\mu_1 \mu_2 \mu_3} 
\int\! \frac{i\, d^{\Dconf} z}{\pi^{{\Dconf}/2}} \,\Delta\left(z-z(\tau_1)\right)\,\Delta\left(z-z(\tau_2)\right)
\,\Delta\left(z-z(\tau_3)\right)\,.
\end{eqnarray}
Here we have indicated with $\Dconf=4-2\epsilon_{UV}$, the dimension
of the configuration space in which the hard diagram lives, and $V^{\mu_1 \mu_2 \mu_3}$ is a differential operator that
represents the three-gluon vertex:
\begin{equation}
  \label{eq:3g-vertex}
  V^{\mu_1 \mu_2 \mu_3} = \eta^{\mu_1 \mu_2} (\partial_1^{\mu_3}-\partial_2^{\mu_3})+
  \eta^{\mu_2 \mu_3} (\partial_2^{\mu_1}-\partial_3^{\mu_1})+
  \eta^{\mu_1 \mu_3} (\partial_3^{\mu_2}-\partial_1^{\mu_2})\,,
\end{equation}
where we have used the notation $\partial_i^\mu \equiv \partial/\partial
z_i^\mu$. Exploiting now the relations
\begin{equation}
  \label{eq:xi-Qi}
  \begin{split}
    x_1-x_2 &= p_1+Q_3\,\qquad z_1-z_2 = \bar \tau_1 p_1+\tau_2 p_2+Q_3\,,\\
    x_2-x_3 &= p_2+Q_1\,\qquad z_2-z_3 = \bar \tau_2 p_2+\tau_3 p_3+Q_1\,,\\
    x_3-x_1 &= p_3+Q_2\,\qquad z_3-z_1 = \bar \tau_3 p_3+\tau_1 p_1+Q_2\,,
  \end{split}
\end{equation}
we can rewrite the tree-gluon vertex in terms of derivatives with
respect to the external momenta $\{Q_1,Q_2,Q_3\}$ only. Introducing
the new differential operator
\begin{equation}
  \label{eq:3g-operator}
  \begin{split}
    \tilde V_{\mu_1 \mu_2 \mu_3} &= \eta_{\mu_1 \mu_2}
    \left(2\frac{\partial}{\partial Q_3^{\mu_3}}-
      \frac{\partial}{\partial Q_1^{\mu_3}}-
      \frac{\partial}{\partial Q_2^{\mu_3}}\right)\\
    &+ \eta_{\mu_2 \mu_3} \left(2\frac{\partial}{\partial
        Q_1^{\mu_1}}- \frac{\partial}{\partial Q_2^{\mu_1}}-
      \frac{\partial}{\partial Q_3^{\mu_1}}\right)\\
    &+\eta_{\mu_1 \mu_3} \left(2\frac{\partial}{\partial
        Q_2^{\mu_2}}-\frac{\partial}{\partial
        Q_1^{\mu_2}}-\frac{\partial}{\partial Q_3^{\mu_2}}\right)\,,
  \end{split}
\end{equation}
and exploiting the fact that the integral in
Eq.~(\ref{eq:hard-contracted}) is finite, we can extract $\tilde
V_{\mu_1 \mu_2 \mu_3}$ from the $z$ integral and write
\begin{equation}
  \label{eq:hard-final}
  \Hard(p_1,p_2,p_3,Q_1,Q_2,Q_3) = p_1^{\mu_1} p_2^{\mu_2}
  p_3^{\mu_3}  \,\tilde V_{\mu_1 \mu_2 \mu_3} 
  \SHard(p_1,p_2,p_3,Q_1,Q_2,Q_3)\,.
\end{equation}
Here we have introduced  a ``scalar hard diagram'', 
defined by:
\begin{equation}
  \label{eq:scalar-hard}
  \SHard(p_1,p_2,p_3,Q_1,Q_2,Q_3) = \int_0^1 \left(\prod_{i=1}^3 d\tau_i\right)
\int\! \frac{i\,d^{\Dconf} z}{\pi^{{\Dconf}/2}} \,\prod_{i=1}^3\Delta\left(z-z(\tau_i)\right)\,.
\end{equation}

\begin{figure}[h]
\begin{center}
  \centering
  \includegraphics[width=.5\textwidth]{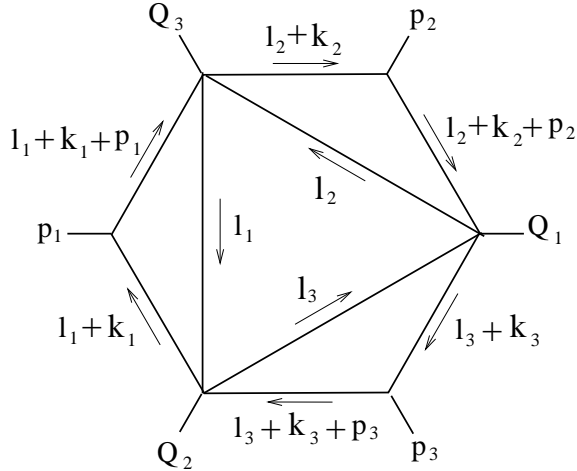}
\end{center}
  \caption{A four-loop scalar hexagon.}
  \label{fig:hexa-4loop}
\end{figure}

\begin{figure}[h] 
\begin{center}
  \includegraphics[width=0.8\textwidth]{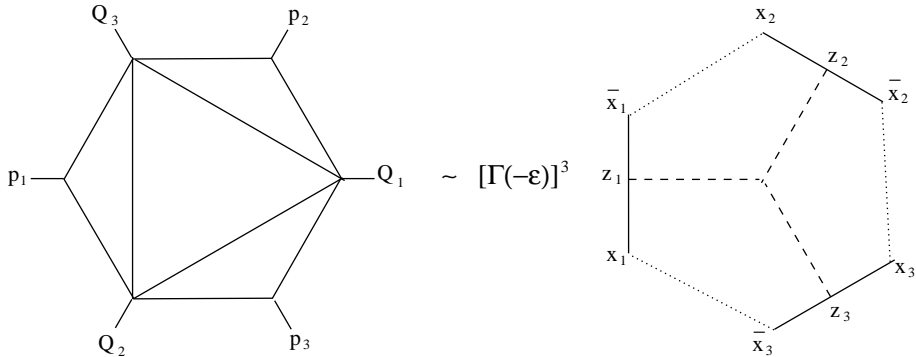}
\end{center}
  \caption{Pictorial representation of the correspondence between a four-loop hexagon and a diagram in configure space.}
  \label{fig:hexa-hard}
\end{figure}

We can turn the scalar hard diagram into a Feynman diagram in momentum
space. First, we introduce three momenta $k_i = z-x_i$ and identify the
position $z$ we are integrating over with the loop momentum
$k=z-x_1=k_1$. Substituting the representation in
Eq.~\eqref{eq:x-triangle} for each propagator $\Delta(z-z_i)$ we
obtain
\begin{multline}
  \label{eq:hard-triangles}
  \SHard(p_1,p_2,p_3,Q_1,Q_2,Q_3) =
  \left(-\frac{\Gamma(1-2\epsilon)}{4\pi^{2+\epsilon}\Gamma(-\epsilon)\Gamma(1-\epsilon)}\right)^3
  \times \\ \times
    \int \frac{i\,d^{\Dconf} k}{\pi^{{\Dconf}/2}}\,
    \prod_{i=1}^3 \left[\Tria(k_i,p_i,1,1,1)\right]^* \,,
\end{multline}
where, as in the previous section, we have taken the complex conjugate
of the triangle to account for the fact that the $\izero$ prescription
of the scalar propagators in $\Tria(k_i,p_i;1,1,1)$ is reversed
with respect to the propagators in configuration space in
Eq.~\eqref{eq:scalar-hard}.  
We further replace each triangle $\Tria(k_i,p_i,1,1,1)$ with its explicit expression of
Eq.~\eqref{eq:triangle-1loop},
and rewrite the scalar hard diagram as
\begin{multline}
  \label{eq:hard-4loop}
  \SHard(p_1,p_2,p_3,Q_1,Q_2,Q_3) =
  \left(-\frac{\Gamma(1-2\epsilon)}{4\pi^{2+\epsilon}\Gamma(-\epsilon)\Gamma(1-\epsilon)}\right)^3
  \times \\ \times
    \int \frac{i\, d^{\Dconf} k}{\pi^{{\Dconf}/2}}\,
  \left(\prod_{i=1}^3 \frac{d^D \ell_i}{i \pi^{\frac{D}{2}}}   
  \frac{1}{\ell_i^2\,(\ell_i+k_i)^2\,
    (\ell_i+k_i+p_i)^2}\right)\,.
\end{multline}
 Notice that the dimension of the $\ell$ integral is
 $D=4-2\epsilon$, as in Eq.~\eqref{eq:triangle-1loop}. The $k$
 integration  is, however, in  $\Dconf=4-2 \epsilon_{UV}=4+2\epsilon$ 
dimensions. We  remark that the diagram is finite for
$\epsilon=-\epsilon_{UV}=0$, and  insensitive to the regularization 
prescription of the various integrations. 
We observe now that, using the definitions of the momenta $k_i=z-x_i$
in terms of the positions $x_i$, and the relations between the
``distances'' $x_{i+1}-x_i$ in Eq.~\eqref{eq:xi-Qi}, we can express
$k_i$ in terms of the light-like momenta $\{p_1,p_2,p_3\}$ and of
three further momenta $\{Q_1,Q_2,Q_3\}$ as follows
\begin{equation}
  \label{eq:hexa-ki}
  k_1 = k\,,\quad
  k_2 = k+p_1+Q_3\,,\quad
  k_3 = k-p_3-Q_2\,.
\end{equation}
We then recognize that the integral in Eq.~\eqref{eq:hard-4loop}
resembles the four-loop hexagon shown in Fig.~\ref{fig:hexa-4loop}:
\begin{equation}
  \label{eq:4loop-hexa}
  \Hexa^{(4)}(p_1,p_2,p_3,Q_1,Q_2,Q_3) = \int \frac{d^D k}{i \pi^{\frac{D}{2}}} 
  \left(\prod_{i=1}^3 \frac{d^D \ell_i}{i \pi^{\frac{D}{2}}}   
  \frac{1}{\ell_i^2\,(\ell_i+k_i)^2\,
    (\ell_i+k_i+p_i)^2}\right)\,.
\end{equation}
We notice that the correspondence is exact only in strictly 
$D=4$ dimensions  because the $k$
integrals in Eq.~\eqref{eq:hard-4loop} and Eq.~\eqref{eq:4loop-hexa}
have to be performed in two different dimensions. 
This correspondence, represented in Fig.~\ref{fig:hexa-hard}, gives promise to  derive 
the finite value  of the  hard diagram  from
the leading  $1/\epsilon^3$ pole of the four-loop hexagon, 
\begin{multline}
  \label{eq:hard-hexa-4d}
    \left.\SHard(p_1,p_2,p_3,Q_1,Q_2,Q_3)\right|_{d=4} =
    \left(\frac{1}{4\pi^2}\right)^3 
    \times \\ \times
    \lim_{\epsilon\to 0}\epsilon^3
    \left.[\Hexa^{(4)}(p_1,p_2,p_3,Q_1,Q_2,Q_3)]^*\right|_{D=4-2\epsilon}\,,
\end{multline}
using factorization properties of collinear singularities. 
The determination of this coefficient would give the analytic answer
for the two-loop vacuum expectation value of a polygon Wilson loop
with an arbitrary number of sides. We leave this calculation for future
work.  

\section{Numerical evaluation  of a  four-loop hexagon Feynman diagram}
\label{sec:hexa-num}

The  dualities that we are  exploring in this paper could  facilitate 
Feynman diagram calculations. We have  not yet explored their  full
potential. However, benefits in computing numerically the multi-loop 
Feynman diagrams which enter  these  dualities  are striking. 
From the Wilson-loop side of the correspondence, the hard diagram is
only difficult to compute analytically and it can be evaluated
numerically rather easily. As  we  have seen in the previous  section, 
this diagram is  dual to a  very complicated four-loop hexagon. 
We  can use the duality to compute  such a complicated  Feynman 
integral. To the best of our knowledge, there has been no  
four-loop Feynman integral with six legs evaluated in the literature.

We start from the explicit expression for the hexagon in
Eq.~\eqref{eq:4loop-hexa} and replace each triangle subgraph with its
representation in terms of a scalar propagator $\Delta(x)$ in
Eq.~\eqref{eq:x-triangle}. Using the correspondence between momenta
and space-time points introduced in Eq.~\eqref{eq:xi-Qi}, we obtain
the following representation:
\begin{multline}
  \label{eq:hexa-hard-shift}
  \mathrm{Hexa}^{(4)}(p_1,p_2,p_3,Q_1,Q_2,Q_3) = 
  \left(-4\pi^{2+\epsilon} \frac{\Gamma(-\epsilon)\,\Gamma(1-\epsilon)}
    {\Gamma(1-2\epsilon)}\right)^3 \times \\ \times
  \int_0^1 \prod_{i=1}^3 d\tau_i
  \int \frac{d^Dz}{i \pi^{\frac{D}{2}}}\, 
  \Delta^*(z)\, \Delta^*(z+z_{12})\, \Delta^*(z-z_{31}) \,,
\end{multline}
where the scalar propagators $\Delta(x)$ are in $4+2\epsilon$
dimensions, while the $z$ integral has to be performed in
$4-2\epsilon$ dimensions.  We now substitute the explicit expression
for $\Delta(x)$, introduce a further Feynman parameterization of the
$z$ integral so as to write the product of propagators as a single
denominator, and finally perform the $z$ integration to get:
\begin{multline}
  \label{eq:hard-for-sector}
  \Hexa^{(4)}(p_1,p_2,p_3,Q_1,Q_2,Q_3) = 
  \left(
    \frac{\Gamma(-\epsilon)\, \Gamma(1-\epsilon)
    }{\Gamma(1-2\epsilon)}\right)^3
  \Gamma(1+4\epsilon) \times \\
  \times
  \int_0^1 \left(\prod_{i=1}^3 d\tau_i\right)
  \int_0^1 \left(\prod_{i=1}^3 d\alpha_i\right) 
  \delta\left(1-\sum_{i=1}^3\alpha_i\right)
  \frac{(\alpha_1\alpha_2\alpha_3)^{\epsilon}}{[-(\alpha_1 \alpha_2 z^2_{12} + \alpha_2 \alpha_3 z^2_{23} +
  \alpha_3 \alpha_1 z^3_{31}+\izero)]^{1+4\epsilon}}\,.
\end{multline}
We obtain an expression for the four-loop hexagon that is a function
of only five Feynman parameters, and that depends on the coordinates
$z_i$ only through the distances $z_{i,i+1}$ introduced in
Eq.~\eqref{eq:xi-Qi}.  This  representation offers  the further  advantage  that  all collinear 
singularities  of the hexagon graph have been captured  as  a divergent  $1/\epsilon^3$ 
prefactor.  The  integral  is finite  in all limits and  can be  expanded directly around  $\epsilon=0$.  

This scalar integral depends on twelve invariants:
\begin{multline}
  \label{eq:hexa-invariants}
\{s_1,\dots,s_{12}\} = 
\{(p_1 p_2),(p_1 p_3), (p_2 p_3), (p_1 Q_3), (p_3 Q_1) , (p_3 Q_2) ,\\
(p_2 Q_3) ,(p_1 Q_2),(p_2 Q_1),Q_1^2,Q_2^2,Q_3^2\}\,.  
\end{multline}
For illustration of numerical results we consider the case in which
all momenta,  $Q_i$,$p_i$ are light-like, so that the final result
depends only on $\{s_1,\dots,s_9\}$.  We cast the result in the form:
\begin{equation}
  \label{eq:hexa-num}
    \Hexa^{(4)}(p_1,p_2,p_3,Q_1,Q_2,Q_3)=\Gamma(1+4\epsilon) \left(\frac{A_{-3}}{\epsilon^3}+\frac{A_{-2}}{\epsilon^2}
    +\frac{A_{-1}}{\epsilon}+A_{0}\right)\,.
\end{equation} 
The results of the numerical integration for two specific sets of invariants are displayed in table~\ref{tab:hexa-num}.
\begin{table}[h]
  \centering
{\footnotesize
\[
  \begin{array}{|c||c|c|c|c|}
    \hline
   \{s_1,\dots,s_9\} & A_{-3} & A_{-2} & A_{-1} & A_{0} \\
     \hline
\{-1,-1,-1,-1,-1,-1,-1,-1,-1\} & -1.04683(4) & 1.2968(1) & 1.7337(2) & 1.9918(4) \\
\hline
\{-0.1,-0.2,-0.3,-0.2,-0.4,-0.6,-1,-0.15,-0.25\}&-2.80209(2) & -7.7878(7) & -4.993(2) & 18.496(6) \\
\hline
  \end{array}
\]
}
 \caption{Numerical evaluation of the four-loop hexagon for different values of the kinematical invariants.}
  \label{tab:hexa-num}
\end{table}

It is intriguing that this four-loop hexagon can be mapped to a
one-loop triangle with non-integers powers of propagators.  From the
above Feynman parameterization we read that
 \begin{eqnarray}
\label{eq:Hexa42Tria}
 &&   \Hexa^{(4)}(p_1,p_2,p_3,Q_1,Q_2,Q_3) = 
\left( 
   \frac{\Gamma(-\epsilon) \Gamma(1-\epsilon)
    }{\Gamma(1+\epsilon)\Gamma(1-2\epsilon)}\right)^3
\int_0^1 d\tau_1 d\tau_2
    d\tau_3 
\nonumber \\
&& \qquad \times
\Tria(\bar \tau_1 p_1+\tau_2 p_2+Q_3,\bar \tau_2 p_2+\tau_3 p_3+Q_1,1+\epsilon,1+\epsilon,1+\epsilon)\,.
\end{eqnarray}
Equivalently, the scalar Wilson-loop ``hard diagram'' through order
${\cal O}(\epsilon^0)$ is also proportional to the same integral over
the one-loop triangle in the right-hand side of the above equation
(with a non-divergent prefactor)~\cite{Anastasiou:2009kn},
  \begin{eqnarray}
\label{eq:Shard2Tria}
  &&   \SHard(p_1,p_2,p_3,Q_1,Q_2,Q_3) = 
\frac{1}{(4 \pi)^3}
\int_0^1 d\tau_1 d\tau_2
    d\tau_3  \nonumber \\ 
&& \qquad 
\times
\Tria(\bar \tau_1 p_1+\tau_2 p_2+Q_3,\bar \tau_2
    p_2+\tau_3 p_3+Q_1,1,1,1) 
+{\cal O}(\epsilon). 
\end{eqnarray} 
In this paper and in Ref.~\cite{Anastasiou:2009kn}  equations like the
above (\ref{eq:Hexa42Tria}-\ref{eq:Shard2Tria}) have been used  for
numerical evaluations of the integrals  on their left-hand side. However, we
believe that  they are  likely to be a  good  starting point for an
analytic evaluation. We remark that the $\Tria$ function is expressed
rather compactly in  terms of Appell $F_4$ functions for arbitrary
powers of  propagators~\cite{Boos:1987bg,Davydychev:1992xr,Anastasiou:1999ui}.

\section{The ``hard'' Wilson-loop diagram as a massive one-loop
  hexagon}
\label{sec:triangle-mass}

\begin{figure}[h]
\begin{center}
 \includegraphics[width=\textwidth]{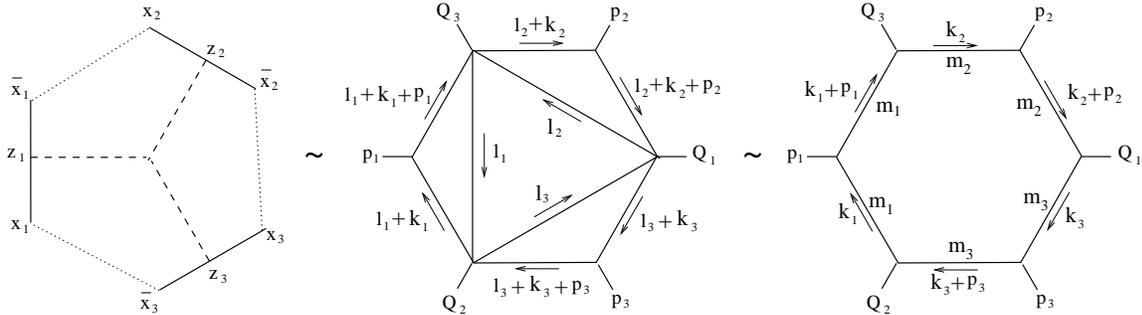}
\end{center}
  \caption{Representation of the mapping of a four-loop hexagon into a one-loop hexagon integrated over internal masses.}
  \label{fig:hexa-mass}
\end{figure}

Wilson-loop integrals can be cast as massive loop Feynman
integrals, integrated over their mass parameters. 
For an arbitrary momentum $q$
\begin{equation}
  \label{eq:mass-integral-n}
  \int_0^{\infty} dm^2 \frac{(m^2)^{-\epsilon}}{\left[(q^2-m^2+\izero)^2\right]^n} = 
  (-1)^n \frac{\Gamma(1-\epsilon)\Gamma(n-1+\epsilon)-1}{\Gamma(n)}{[-q^2-\izero]^{n-1+\epsilon}}. 
\end{equation}
For $n=2$, we obtain a representation of the propagator in
configuration space:
\begin{equation}
  \label{eq:scalar-prop-mass}
  \Delta^{*}(q) =\frac{1}{4\pi^{2+\epsilon}} \frac{\Gamma(1+\epsilon)}
{(-q^2-\izero)^{1+\epsilon}}\, = \frac{1}{4 \pi^{2+\epsilon}
  \Gamma(1-\epsilon)} \int_0^\infty dm^2 \frac{ (m^2)^{-\epsilon}}{\left[q^2-m^2+\izero\right]^2}.
\end{equation}
For a propagator connecting a fixed point $x_0=-k$ and a light-like
line segment $x(\tau) = \tau p$, we have $q=k+\tau p$. Performing the
$\tau$ integral we have
\begin{equation}
 \int_0^1 d\tau \Delta^{*}(k+\tau p) =
\frac{1}{4 \pi^{2+\epsilon}
  \Gamma(1-\epsilon)} \int_0^\infty dm^2 \frac{ (m^2)^{-\epsilon}}{
\left[k^2-m^2+\izero\right]
\left[(k+p)^2-m^2+\izero\right]
}\,.
\end{equation}
The same relation  (up to prefactors) holds for the dual one-loop
triangle. 
We find
\begin{equation}
{\rm Tria}(k,p;1,1,1) = \frac{\Gamma(-\epsilon)}{\Gamma(1-2\epsilon)} 
\int_0^\infty dm^2 \frac{(m^2)^{-\epsilon}}{(k^2-m^2+\izero)
  \left[(k+\tau p)^2 - m^2 + \izero \right]}\,.
\end{equation}
\begin{figure}[h]
\begin{center}
  \includegraphics[width=0.85\textwidth]{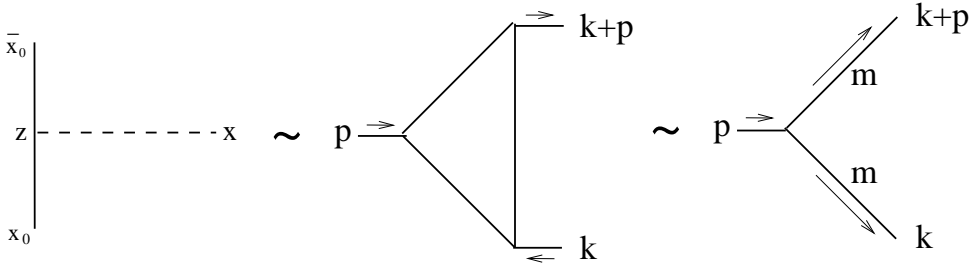}
\end{center}
 \caption{Pictorial representation of the one-loop triangle as an integral over
    masses.}
\label{fig:triangle-mass}
\end{figure}
This correspondence, pictorially represented in
Fig.~\ref{fig:triangle-mass}, leads to a different representation for
the four-loop hexagon considered in Section~\ref{sec:hexa-4loop} and
displayed in Fig.~\ref{fig:hexa-4loop} and, equivalently, for the
``hard'' Wilson-loop diagram.  We find:
\begin{multline}
  \label{eq:hexa-mass}
  \Hexa^{(4)}(p_1,p_2,p_3,Q_1,Q_2,Q_3) =
  \left(
    \frac{\Gamma(-\epsilon)}{\Gamma(1-2\epsilon)}\right)^3
\times      \\
  \times \int \frac{d^4k}{i \pi^2} \int_0^{\infty} \prod_{i=1}^3 dm_i^2
  \frac{(m_i^2)^{-\epsilon}}{(k_i^2-m_i^2+ \izero)[(k_i+p_i)^2-m_i^2+\izero]}\,,
\end{multline}
where $k_1=k$, $k_2=k+p_1+Q_3$ and $k_3=k-Q_2-p_3$.   
Similarly, the  Wilson-loop hard diagram can be written as (see Fig.~\ref{fig:hexa-mass})
\begin{multline}
  \label{eq:shard-mass}
  \SHard^{*}(p_1,p_2,p_3,Q_1,Q_2,Q_3) =
  \left( -4 \pi^{2+\epsilon} \Gamma(1-\epsilon)\right)^3
    \times \\ 
  \times \int \frac{d^4k}{i \pi^2} \int_0^{\infty} \prod_{i=1}^3 dm_i^2
  \frac{(m_i^2)^{-\epsilon}}{(k_i^2-m_i^2+ \izero)[(k_i+p_i)^2-m_i^2+\izero]}\,.
\end{multline}
The four-dimensional massive one-loop hexagon can be evaluated
analytically after reducing to box master integrals.  The analytic
integration over the masses $m_i$ is a formidable task.  However, it
holds promise for achieving an analytic evaluation of the hard diagram
for a Wilson loop with an arbitrary number of sides.

\section{Non-planar Feynman integrals}
\label{sec:non-planar}

An intriguing feature of Eq.~(\ref{eq:ebox-final}) is that an
``easy box'' is cast as a single scalar propagator raised to a
dimension-dependent power.  Such boxes can be subgraphs of more
complicated higher loop non-planar diagrams. Eq.~(\ref{eq:ebox-final})
can be then utilized to cast these non-planar diagrams as planar
diagrams with one loop less, integrated over a range of linear
combinations for their external momenta.

\begin{figure}[h]
 \begin{center}
  \includegraphics[width=0.85\textwidth]{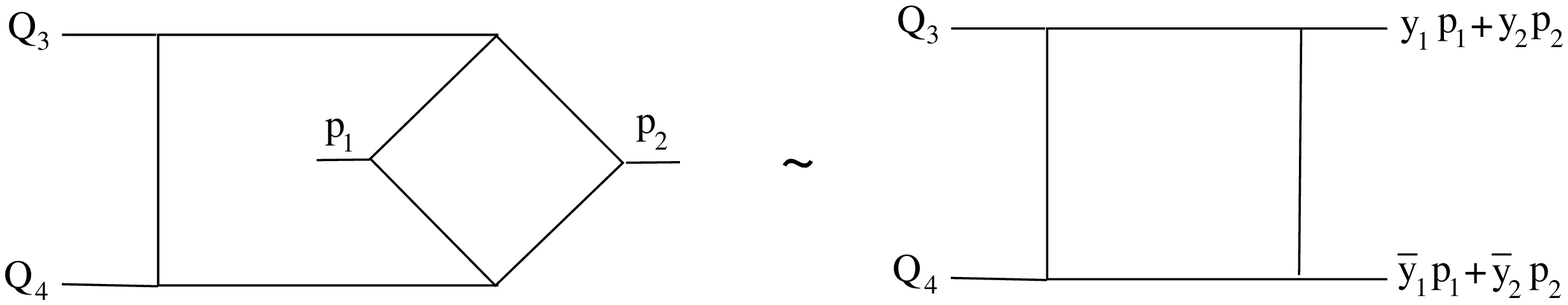}
\end{center}
  \caption{Cross box diagrams contributing to $2\to 2$ processes.}
  \label{fig:cross-box-2loop}
\end{figure}
The two-loop non-planar box integral ${\rm XBox}$ at the left of Fig.~\ref{fig:cross-box-2loop}
has   a one-loop ``easy-box'' subgraph.    
Applying Eq.~(\ref{eq:ebox-final}), we obtain a representation as an integral
\begin{multline}
  \label{eq:xbox-massless}
  \XBox = 2 \, (-1)^{-\epsilon}\,
  \Gamma(2+\epsilon)\,\frac{\Gamma(-\epsilon)\Gamma(1-\epsilon)}{\Gamma(1-2\epsilon)}\,
\int_0^1 dy_1 \! \int_0^1 dy_2\, \times \\ \times 
{\cal F}\left( (p_1+p_2)^2,  (y_1p_1+y_2p_2+Q_3)^2,  (y_1p_1+y_2p_2)^2, (\bar{y}_1p_1+\bar{y}_2p_2+Q_3)^2  , Q_3^2\right),   
\end{multline}
over a  box function
\begin{multline}
{\cal F}\left( (p_1+p_2)^2,  (y_1p_1+y_2p_2+Q_3)^2,  (y_1p_1+y_2p_2)^2, (\bar{y}_1p_1+\bar{y}_2p_2+Q_3)^2, Q_3^2 \right)  = \\
\int \frac{d^Dk}{ i \pi^{\frac{D}{2}}} \frac{1}{k^2 \left[(k+y_1p_1+y_2p_2)^2\right]^{2+\epsilon}(k+p_1+p_2)^2 (k+p_1+p_2+Q_3)^2},
\end{multline}
with one of the propagators raised to a non-integer power. 

We observe that this representation requires at most five Feynman
parameters (three for the box-function in the integrand and
$y_1,y_2$).  A naive Feynman representation would yield a
six-dimensional integral.  One-loop box integrals have been studied
extensively in the literature.  In Eq.~(\ref{eq:xbox-massless}) it is
required a box integral with a non-integer power for one of the
propagators.  For $Q_3^2=Q_4^2$ this is known to be a sum of four
$F_4$ hypergeometric
functions~\cite{Anastasiou:1999ui,Anastasiou:1999cx} with well studied
analytic continuation properties and asymptotic limits in the
mathematical literature. 

For a direct numerical evaluation with the method  of sector decomposition~\cite{Binoth:2000ps} 
this  Feynman representation  is better suited requiring  less  than half  the number of 
sectors of a naive parameterization.  Applying the non-linear transformations of 
Ref.~\cite{Anastasiou:2010pw} to factorize  infrared  singularities is  also a much simpler task with our  parameterization.   
Non-planar box diagrams with light-like external legs pose an
additional difficulty for their evaluation due to not having a
Euclidean region when the Mandelstam variables $s =(p_1+p_2)^2$,
$t=(p_2+Q_3)^2$, $u=(Q_3+p_1)^2$ are consistent with momentum
conservation:
\begin{equation}
Q_3^2=Q_4^2= s+t+u=0.
\end{equation}
The on-shell limit of an external leg does not commute with the $\epsilon = 0$ of  the dimensional regulator due to the emergence  of new 
collinear divergences.  It is easy to obtain a Feynman representation with a Euclidean region 
for generic  $s,t,u$ from   Eq.~(\ref{eq:xbox-massless}) which possesses the  same  infrared singularities  as for  $s,t,u=-s-t$. This is,
\begin{multline}
  \label{eq:xbox-massless1}
  \XBox = 2 \, (-1)^{-\epsilon}\,
  \Gamma(2+\epsilon)\,\frac{\Gamma(-\epsilon)\Gamma(1-\epsilon)}{\Gamma(1-2\epsilon)}\,
\int_0^1 dy_1 \! \int_0^1 dy_2\, \times \\ \times 
{\cal F}\left( s,  y_1 \bar{y}_2 t + y_2 \bar{y}_1 u , y_1y_2 s, \bar{y}_1\bar{y_2}s,0 \right),   
\end{multline}
where  we have used momentum conservation only for the second argument of the ${\cal F}$ box function.  

Using the Mellin-Barnes representation of the one-loop box with two
off-shell legs, and introducing one additional Mellin-Barnes
integration, we can integrate out all Feynman parameters.  We obtain
\begin{multline}
  \label{eq:xbox-MB-final}
  \XBox(p_1,p_2,Q_3;0,0,0) = -2 \, \,
  \frac{\Gamma(-\epsilon)\Gamma(1-\epsilon)}{\Gamma(-1-3\epsilon)\Gamma(1-2\epsilon)}  \int_{-i\infty}^{i\infty} \left(\prod_{i=1}^4
    \frac{d\xi_i}{2\pi i} \Gamma(-\xi_i)\right)
  \times \\ \times
  \Gamma(3+2\epsilon+\xi_{1234})\Gamma(2+\epsilon+\xi_{1234})
  \Gamma(-2-2\epsilon-\xi_{134})\Gamma(-2-2\epsilon-\xi_{234}) 
  \times \\ \times
  \frac{\Gamma(1+\xi_{34})\Gamma(1+\xi_{13}) \Gamma(1+\xi_{14}) \Gamma(1+\xi_{23}) \Gamma(1+\xi_{24}) }{\Gamma^2(2+\xi_{1234})} (-s)^{-3-2\epsilon-\xi_{34}} (-t)^{\xi_3} (-u)^{\xi_4}\,,
\end{multline}
which is the representation obtain in ref.~\cite{Tausk:1999vh}.

Similarly simplified Feynman parameterizations  of non-planar  diagrams  with ``easy-box'' insertions  can be also 
obtained for integrals with higher number of loops or legs. 

\section{Conclusions}
We have observed dualities among diagrams which enter the calculation
of the vacuum expectation value of  Wilson loops and scalar Feynman
integrals.  These can be pictured as:
 \begin{center}
  \includegraphics[width=0.85\textwidth]{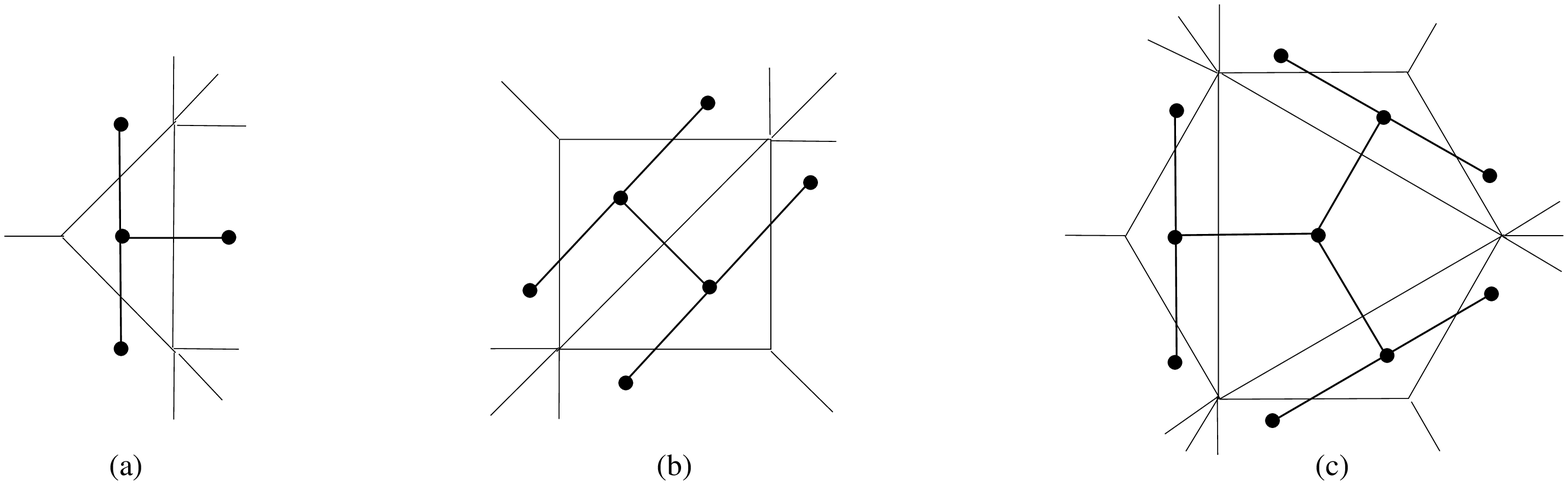}
\end{center}
where the meaning of each diagram is:
\begin{enumerate}
\item[(a)] a one-loop triangle is dual to a scalar propagator in
  configuration space attached to a Wilson line;
\item[(b)] a two-loop diagonal box (and a one-loop easy box) are dual to
  a scalar propagator connecting two Wilson lines, in fact a one-loop
  Wilson-loop with an arbitrary number of edges;
\item[(c)] a four-loop hexagon is dual to a three-boson vertex
  connected via scalar propagators to three Wilson lines, which is the
  so-called two-loop ``hard diagram''.
\end{enumerate}
A remarkable feature of the representations we have obtained is that
the Wilson-line diagram appears always as a factor multiplying a
singularity of the corresponding scalar multi-loop diagram.  This
factorization property has two main advantages. On one hand, if one
wishes to exploit a duality to compute an unknown Wilson-loop diagram,
it is in general enough to extract the coefficient of the leading
singularity of the corresponding scalar integral, which is far easier
than fully computing the integral itself. On the other hand, given a
scalar multi-loop integral, the corresponding Wilson-loop diagram
provides automatically a nice Feynman parameterization for
it. In particular, given the fact that part of the singularities are
already extracted and that Wilson loops are generally simple to
compute numerically, one can use the duality to tackle calculations
that would appear impossible at first sight. For illustration of the
potential our observations, we compute for the first time in the
literature a four-loop hexagon integral, which is dual to the
Wilson-loop ``hard diagram''.

The relations we have found concern planar diagrams only. However,
they can be also used when a planar diagram is part of a larger
non-planar diagram. For instance, one can use the duality between the
one-loop easy box and a scalar propagator to derive a simple Feynman
parameterization for the two-loop cross-box, yielding a better
starting point for its numerical
evaluation~\cite{Anastasiou:2010pw}. The very same parameterization
simplifies the calculation of the cross-box using Mellin-Barnes
techniques.

To conclude, the dualities we have found look very promising, and can
lead to significant simplifications in the calculation of high-loop
integrals appearing both in QCD and in other theories like
$\mathcal{N}=4$ Super Yang-Mills or $\mathcal{N}=8$ Supergravity.  We
are looking forward to further investigations on these dualities in
the future.

\paragraph{Acknowledgments.} 
We thank Andreas Brandhuber, Lance  Dixon, Vittorio del Duca,  Claude Duhr,  
Thomas Gehrmann, Gregory Korchemsky, Eric Laenen, Lorenzo Magnea, 
Volodya Smirnov,  Bas Tausk  and Gang Yang for useful discussions.  
We thank especially Bas Tausk for communicating his  insight  on the  diagonal-box/one-loop box duality 
in the past.  This work  is  supported by the ERC Starting Grant project IterQCD.



\begin{thebibliography}{99}

\bibitem{Maldacena:1997re}
  J.~M.~Maldacena,
  {\it The large $N$ limit of superconformal field theories and supergravity},
  Adv.\ Theor.\ Math.\ Phys.\  {\bf 2} (1998) 231
  [Int.\ J.\ Theor.\ Phys.\  {\bf 38} (1999) 1113]
  [\hepth{9711200}].

\bibitem{Minahan:2002ve}
  J.~A.~Minahan and K.~Zarembo,
  {\it The Bethe-ansatz for N = 4 super Yang-Mills},
  JHEP {\bf 0303} (2003) 013
  [ hepth{0212208}].

\bibitem{Beisert:2003ea}
  N.~Beisert, S.~Frolov, M.~Staudacher and A.~A.~Tseytlin,
  {Precision spectroscopy of AdS/CFT},'
  JHEP {\bf 0310} (2003) 037
  [\hepth{0308117}], and references therein.

\bibitem{Beisert:2005fw}
  N.~Beisert and M.~Staudacher,
  {\it Long-range $PSU(2,2|4)$ Bethe ansaetze for gauge theory and strings},
  Nucl.\ Phys.\  B {\bf 727} (2005) 1
  [\hepth{0504190}], and references therein.

\bibitem{Anastasiou:2003kj}
  C.~Anastasiou, Z.~Bern, L.~J.~Dixon and D.~A.~Kosower,
  {\it Planar amplitudes in maximally supersymmetric Yang-Mills theory},
  Phys.\ Rev.\ Lett.\  {\bf 91} (2003) 251602
  [\hepth{0309040}].

\bibitem{Bern:2005iz}
  Z.~Bern, L.~J.~Dixon and V.~A.~Smirnov,
  {\it Iteration of planar amplitudes in maximally supersymmetric Yang-Mills
  theory at three loops and beyond},
  Phys.\ Rev.\  D {\bf 72} (2005) 085001,
  [\hepth{0505205}].

\bibitem{Cachazo:2006tj}
  F.~Cachazo, M.~Spradlin and A.~Volovich,
  {\it Iterative structure within the five-particle two-loop amplitude},
  Phys.\ Rev.\  D {\bf 74} (2006) 045020
  [\hepth{0602228}].

\bibitem{Bern:2006vw}
  Z.~Bern, M.~Czakon, D.~A.~Kosower, R.~Roiban and V.~A.~Smirnov,
  {\it Two-loop iteration of five-point {\cal N} = 4 super-Yang-Mills amplitudes},
  Phys.\ Rev.\ Lett.\  {\bf 97} (2006) 181601
  [\hepth{0604074}].

\bibitem{Bern:2008ap}
  Z.~Bern, L.~J.~Dixon, D.~A.~Kosower, R.~Roiban, M.~Spradlin, C.~Vergu
  and A.~Volovich,
  {\it The Two-Loop Six-Gluon MHV Amplitude in Maximally Supersymmetric
  Yang-Mills
  Theory}, Phys.\ Rev.\  D {\bf 78} (2008) 045007
  [{\tt 0803.1465[hep-th]}].


\bibitem{Alday:2007hr}
  L.~F.~Alday and J.~Maldacena,
  {\it Gluon scattering amplitudes at strong coupling},
  JHEP {\bf 0706} (2007) 064
  [{\tt 0705.0303[hep-th]}].

\bibitem{Drummond:2007aua}
  G.~P.~Korchemsky, J.~M.~Drummond and E.~Sokatchev,
  {\it Conformal properties of four-gluon planar amplitudes and Wilson loops},
  Nucl.\ Phys.\  B {\bf 795} (2008) 385
  [{\tt 0707.0243[hep-th]}].

\bibitem{Brandhuber:2007yx}
  A.~Brandhuber, P.~Heslop and G.~Travaglini,
  {\it MHV Amplitudes in N=4 Super Yang-Mills and Wilson Loops},
  Nucl.\ Phys.\  B {\bf 794} (2008) 231
  [{\tt 0707.1153[hep-th]}].

\bibitem{Drummond:2008aq}
  J.~M.~Drummond, J.~Henn, G.~P.~Korchemsky and E.~Sokatchev,
  {\it Hexagon Wilson loop = six-gluon MHV amplitude},
  Nucl.\ Phys.\  B {\bf 815} (2009) 142
  [{\tt 0803.1466[hep-th]}].

\bibitem{Brandhuber:2010bj}
  A.~Brandhuber, P.~Heslop, P.~Katsaroumpas, D.~Nguyen, B.~Spence, M.~Spradlin and G.~Travaglini,
  {\it A Surprise in the Amplitude/Wilson Loop Duality},
  JHEP {\bf 1007} (2010) 080
  [{\tt 1004.2855[hep-th]}].



\bibitem{DelDuca:2009au}
  V.~Del Duca, C.~Duhr and V.~A.~Smirnov,
  {\it An Analytic Result for the Two-Loop Hexagon Wilson Loop in
    {\cal N} = 4 SYM},
  JHEP {\bf 1003} (2010) 099
  [{\tt 0911.5332[hep-ph]}].

\bibitem{DelDuca:2010zg}
  V.~Del Duca, C.~Duhr and V.~A.~Smirnov,
  {\it The Two-Loop Hexagon Wilson Loop in {\cal N} = 4 SYM}
  JHEP {\bf 1005} (2010) 084
 [{\tt 1003.1702[hep-th]}].



\bibitem{Anastasiou:2009kn}
  C.~Anastasiou, A.~Brandhuber, P.~Heslop, V.~V.~Khoze, B.~Spence and
  G.~Travaglini,
  {\it Two-Loop Polygon Wilson Loops in {\cal N}=4 SYM,}
  JHEP {\bf 0905} (2009) 115
  [{\tt 0902.2245[hep-th]}].

\bibitem{Green:1982sw}
  M.~B.~Green, J.~H.~Schwarz and L.~Brink,
  {\it N=4 Yang-Mills And N=8 Supergravity As Limits Of String Theories},
  Nucl.\ Phys.\  B {\bf 198} (1982) 474.

\bibitem{Bern:1997nh}
  Z.~Bern, J.~S.~Rozowsky and B.~Yan,
  {\it Two-loop four-gluon amplitudes in $\mathcal{N} = 4$ super-Yang-Mills},
  Phys.\ Lett.\  B {\bf 401} (1997) 273
  [\hepph{9702424}].


\bibitem{Drummond:2007au}
  J.~M.~Drummond, J.~Henn, G.~P.~Korchemsky and E.~Sokatchev,
  {\it Conformal Ward identities for Wilson loops and a test of the duality with
  gluon amplitudes},
Nucl.\ Phys.\  B {\bf 826} (2010) 337
[{\tt 0712.1223[hep-th]}].

\bibitem{DelDuca:2010zp}
  V.~Del Duca, C.~Duhr and V.~A.~Smirnov,
  {\it A Two-Loop Octagon Wilson Loop in {\cal N} = 4 SYM},
  JHEP {\bf 1009} (2010) 015 
  [{\tt 1006.4127[hep-th]}].

\bibitem{Gorsky:2009nv}
  A.~Gorsky and A.~Zhiboedov,
  {\it One-loop derivation of the Wilson polygon - MHV amplitude duality},
  J.\ Phys.\ A  {\bf 42} (2009) 355214
  [{\tt 0904.0381[hep-th]}].

\bibitem{Gorsky:2009dr}
  A.~Gorsky and A.~Zhiboedov,
  {\it Aspects of the $\mathcal{N}=4$ SYM amplitude -- Wilson polygon duality},
  Nucl.\ Phys.\  B {\bf 835} (2010) 343
  [{\tt 0911.3626[hep-th]}].

\bibitem{Tausk:1999vh}
  J.~B.~Tausk,
  {\it Non-planar massless two-loop Feynman diagrams with four on-shell legs},
  Phys.\ Lett.\  B {\bf 469} (1999) 225
  [\hepph{9909506}].

\bibitem{Kramer:1986sr}
  G.~Kramer and B.~Lampe,
  {\it Integrals for two loop calculations in massless QCD},
  J.\ Math.\ Phys.\  {\bf 28} (1987) 945.

\bibitem{Boos:1987bg}
  E.~E.~Boos, A.~I.~Davydychev,
 {\it A Method Of The Evaluation Of The Vertex Type Feynman Integrals},                                                                             
 Moscow Univ.\ Phys.\ Bull.\  {\bf 42N3 } (1987)  6-10.
                      
\bibitem{Davydychev:1992xr}
  A.~I.~Davydychev,
 {\it Recursive algorithm of evaluating vertex type Feynman integrals},
  J.\ Phys.\ A {\bf A25 } (1992)  5587-5596.

\bibitem{Anastasiou:1999ui}
  C.~Anastasiou, E.~W.~N.~Glover and C.~Oleari,
  {\it Scalar One-loop integrals using the negative-dimension approach},
  Nucl.\ Phys.\  B {\bf 572}, 307 (2000)
  [\hepph{9907494}].

\bibitem{Anastasiou:1999cx}
  C.~Anastasiou, E.~W.~N.~Glover and C.~Oleari,
  {\it Application of the negative-dimension approach to massless scalar box
  integrals},
  Nucl.\ Phys.\  B {\bf 565}, 445 (2000)
  [\hepph{9907523}].

\bibitem{Binoth:2000ps}
  T.~Binoth and G.~Heinrich,
  {\it An automatized algorithm to compute infrared divergent multi-loop
  integrals},
  Nucl.\ Phys.\  B {\bf 585} (2000) 741
  [\hepph{0004013}].


\bibitem{Anastasiou:2010pw}
  C.~Anastasiou, F.~Herzog, A.~Lazopoulos,
  {\it On the factorization of overlapping singularities at NNLO},
  {\tt 1011.4867[hep-ph]}.

\end{thebibliography}
\end{document}